\documentclass[aps,prx,twocolumn,amsmath,amssymb,superscriptaddress,floatfix]{revtex4-1}
\usepackage{amsmath}
\usepackage{amssymb}
\usepackage{tikzsymbols}
\usepackage{amsfonts}
\usepackage{listings}
\usepackage{enumerate}
\usepackage{latexsym}
\usepackage{bm}
\usepackage{graphicx}
\usepackage{subfigure}
\usepackage{braket}
\usepackage[pdftex,colorlinks=false]{hyperref}
\usepackage{xcolor}
\usepackage{verbatim}

\usepackage{times}

\newcommand{\beq}{\begin{equation}}
\newcommand{\eneq}{\end{equation}}

\usepackage{xr}

\begin{document}

\title{Majorana Quantum Lego, a Route Towards Fracton Matter}

\author{Yizhi You \Summertree}
\affiliation{\mbox{Princeton Center for Theoretical Science, Princeton University, Princeton NJ 08544, USA}}

\author{Felix von Oppen \Snowman}
\affiliation{\mbox{Dahlem Center for Complex Quantum Systems and Fachbereich Physik, Freie Universit\"at Berlin, 14195 Berlin, Germany}}

\affiliation{Institute of Quantum Information and Matter, California Institute of Technology, Pasadena, California 91125, USA}

\date{\today}

\begin{abstract}
Fracton topological phases host fractionalized topological quasiparticles with restricted mobility, with promising applications to fault-tolerant quantum computation. While a variety of exactly solvable fracton models have been proposed, there is a need for platforms to realize them experimentally. We show that a rich set of fracton phases emerges in interacting Majorana band models whose building blocks are within experimental reach. Specifically, our Majorana constructions overcome a principal obstacle, namely the implementation of the complicated spin cluster interactions underlying fracton stabilizer codes. The basic building blocks of the proposed constructions include Coulomb blockaded Majorana islands and weak inter-island Majorana hybridizations. This setting produces a wide variety of fracton states and promises numerous opportunities for probing and controlling fracton phases experimentally. Our approach also reveals the relation between fracton phases and Majorana fermion codes and further generates a hierarchy of fracton spin liquids.

\end{abstract}

\maketitle 

Searching for and exploring exotic phases of matter is a principal goal of condensed matter physics. In the presence of strong interactions, quantum many-body systems composed of a limited number of elementary particles assume a remarkable variety of exotic phases whose low-energy degrees of freedom are much richer than suggested by their constituents. Prominent examples of such emergent quantum phases are topological phases, whose quasiparticle excitations carry fractional quantum numbers and obey anyonic statistics \cite{wen1990topological,willett1987observation}. The low-energy properties of these topologically ordered states are characterized by topological quantum field theories (TQFT) \cite{wen1992classification,dijkgraaf1990topological}.

Recently distinct long-range entangled states, transcending the TQFT paradigm and termed fracton phases, have been discovered and intensively studied in exactly solvable lattice models \cite{Haah2011-ny,Halasz2017-ov,Vijay2016-dr,Vijay2015-jj,Chamon2005-fc}.
Fracton topological order shares many features with topological order, including nontrivial braiding statistics and symmetry fractionalization. At the same time, fracton phases have a subextensive ground-state degeneracy depending on system size in addition to lattice topology, and  quasiparticles with restricted mobility, moving within lower-dimensional manifolds such as planes, lines, or fractals. The subdimensional nature of fracton excitations gives rise to unconventional features including glassiness and subdiffusive dynamics \cite{Chamon2005-fc,prem2017glassy}. The restricted  quasiparticle mobility makes fracton stabilizer codes interesting for quantum-memory and quantum-computation applications \cite{kitaev2003fault,Haah2011-ny}.

While theoretical aspects of fracton phases have been intensively explored via quantum stabilizer codes as well as higher rank gauge theories \cite{pretko2017fracton,ma2018fracton,bulmash2018higgs,Slagle2017-gk}, their physical realization remains a key challenge of condensed matter physics and quantum information science \cite{slagle2017fracton,hsieh2017fractons,Halasz2017-ov}. The principal obstacle is that models exhibiting fracton physics tend to involve rather complicated spin cluster interactions. This raises the question whether and how such exotic fracton states emerge in models with more physical ingredients and interactions. These might then be amenable to experimental implementation, and assuming tunable tunable interaction parameters, allow for controlling and manipulating fracton phases and excitations, e.g., for quantum computing purposes.

Here, we show that many known fracton stabilizer codes can be obtained from Majorana band models with strong onsite interactions. There is currently a major push to develop the required technology for realizing such models in the context of Majorana-based quantum computing \cite{Lutchyn2018rev}, and our work shows how to leverage this budding technology for investigating fracton phases of matter. Our constructions include fracton phases of both flavors, referred to as type-I and type-II fracton codes. While excitations of the former are created by line- or membrane operators, excitations of type-II fractons are generated by operators with fractal support \footnote{Here, we define type-II fracton codes as having excitations which are created by operators with fractal support. We do not require the stricter condition that {\em all} excitations are generated by such fractal operators.}.

The principal ingredients of our constructions are Majorana hybridization as well as local interactions which fix local fermion parities \cite{vijay2015majorana,2015arXiv150907135B,Landau2016SC,Sagi2018,Wille2018,Thomson2018}. These interactions can be implemented using Majorana islands, also referred to as Majorana Cooper pair boxes, which underlie current designs for Majorana-based topological qubits \cite{Lutchyn2018rev,vijay2015majorana,Karzig2017scale,Plugge2017box}. Each island contains some number of Majoranas, e.g., at the ends of semiconductor wires proximity coupled to a superconductor \cite{Lutchyn2010,Oreg2010,Lutchyn2018rev}. The island's charging energy fixes its fermion parity, corresponding to a multi-Majorana interaction. In order to realize fracton phases, the fermion parity typically has to be fixed for overlapping sets of Majoranas. We show how such interactions can be realized by judiciously hybridizing Majoranas from multiple Majorana islands.

Our construction opens new avenues in the study of fracton phases from both conceptual and applied perspectives. On the theoretical side, our Majorana-based setups for fracton models imply that fractonic phases of matter can emerge from strongly interacting one-dimensional $p$-wave superconductors. Moreover, they expose the close relation between 3D Majorana fermion codes \cite{Bravyi2011-fl} and fracton codes, and further generate a hierarchy of 3D SO(N) invariant spin liquids with fractonic behavior. Analogous to Ref.~\cite{shirley2018fractional},
this correspondence also establishes a foliation-based equivalence relation between bosonic and fermionic fracton states. 
On the experimental side, our Majorana-based models not only provide a possible route towards realizing fracton phases of matter, but might also give access to interesting observables of fracton phases and their phase transitions. In particular, the Hamiltonian would be amenable to tuning it away from the stabilizer limit, opening a path towards exploring confinement and disorder effects on fracton matter.

\begin{figure}[t]
  \centering
    \includegraphics[width=0.5\textwidth]{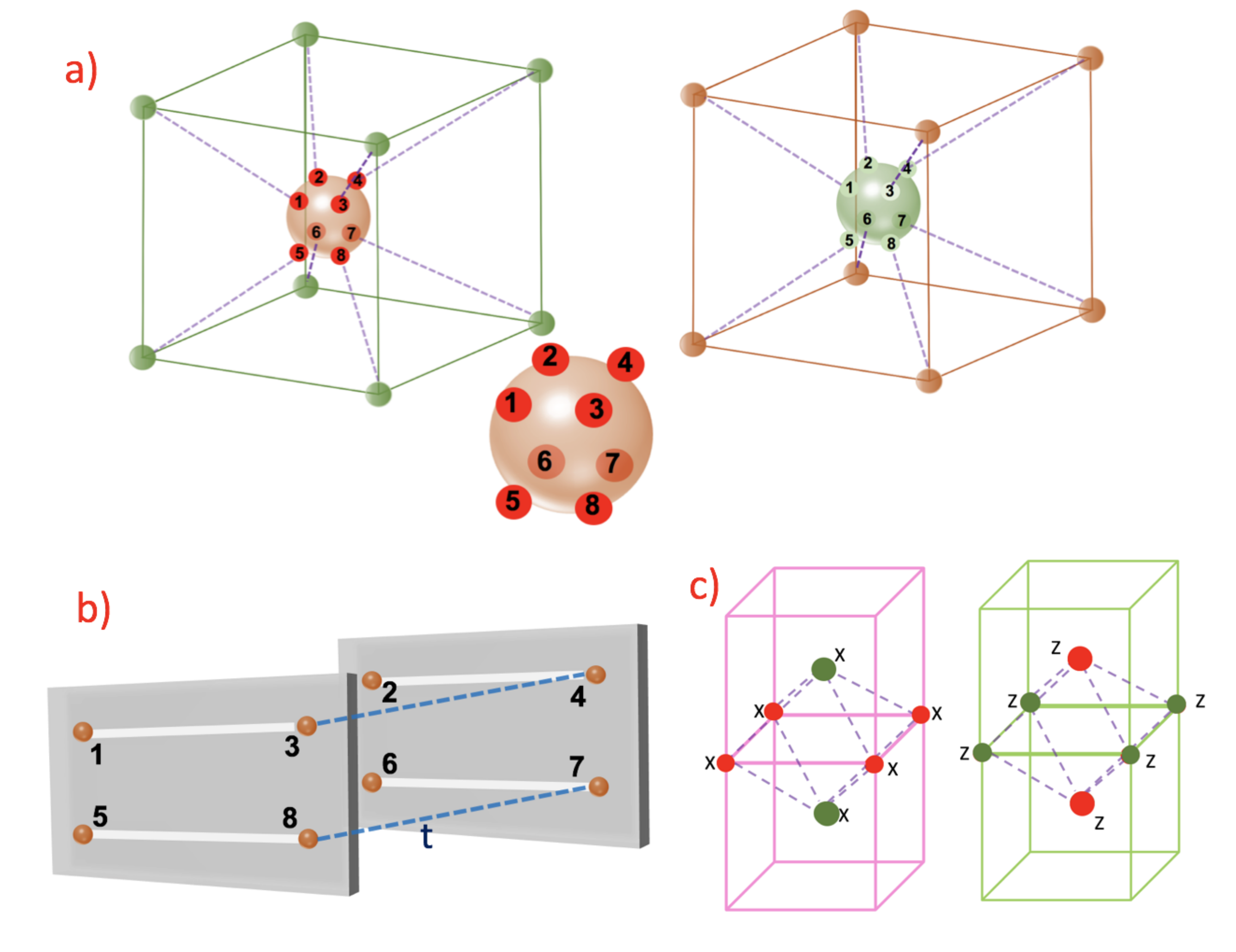} 
  \caption{Construction underlying the Majorana model for the planon-lineon code. a) Body-centered cubic lattice with eight Majorana zero modes on all corner (green) and center (red) sites. Each Majorana pairs with a nearby partner as illustrated by dashed lines. b) Setup for realizing the Majorana quartet interaction in Eq.~(\ref{int}). Two Majorana quartets (red dots) are placed on floating superconducting islands, fixing the corresponding fermion parities by virtue of the charging energy. The third Majorana quartet interaction in Eq.~(\ref{int}) is generated by the pairwise Majorana hybridizations indicated by the dashed lines. 
  c) The two types of octahedral cells which support the stabilizers of the planon-lineon code resulting from the strong-coupling projection of the Majorana model in Eq.~(\ref{H}).} 
    \label{cx}
\end{figure}

\begin{center}
\textbf{
The planon-lineon code}
\end{center}
We begin with a topological superconductor on a body-centered cubic lattice. Each site contains eight Majoranas $\gamma^1, \ldots,\gamma^8$ which are each hybridized with a Majorana on a nearest-neighbor site as shown in Fig.~\ref{cx}. Thus, the topological superconductor has the Hamiltonian \begin{align} 
H=-it' \sum_{\langle i,j\rangle} & (\gamma_i^1 \gamma_{j}^7+ \gamma_i^2 \gamma_{j}^8+ \gamma_i^4 \gamma_{j}^5+ \gamma_i^3 \gamma_{j}^6)
\label{H}
\end{align}
and can be thought of as built from crossing one-dimensional Kitaev chains along the $(\pm1,\pm1,1)$ directions. 

We now consider onsite interactions which couple quartets of Majoranas,  
 \begin{align} 
&H_{\rm int}=U(\gamma^1_i \gamma^3_i \gamma^8_i \gamma^5_i+\gamma^3_i \gamma^4_i \gamma^7_i \gamma^8_i+\gamma^4_i \gamma^2_i \gamma^6_i \gamma^7_i)
\label{int}
\end{align}
These interactions suppress hopping of single Majoranas between sites. In the strong-$U$ limit, they project each site into the $\gamma^1_i \gamma^3_i \gamma^8_i \gamma^5_i=\gamma^3_i \gamma^4_i \gamma^7_i \gamma^8_i=\gamma^4_i \gamma^2_i \gamma^6_i \gamma^7_i=-1$ subspace. The product of the three parity constraints also implies $\gamma^2_i \gamma^1_i \gamma^5_i \gamma^6_i=-1$, constraining the Majorana quartets associated with the four vertical faces of the red cube in Fig.~\ref{cx}. 

Under these parity constraints, each site retains a single spin-${1}/{2}$ degree of freedom. Indeed, the product of the parities of any two opposing faces is constrained to unity, including the top and bottom faces, $(\gamma^1_i \gamma^2_i \gamma^4_i \gamma^3_i)(\gamma^8_i \gamma^7_i \gamma^6_i \gamma^5_i)=1$. We can then choose the (identical) parities of the top and bottom faces as  
the Pauli-$Z$ operator $\sigma_i^z=\gamma^1_i \gamma^2_i \gamma^4_i \gamma^3_i$ and the product of two Majoranas associated with any vertical edge as the Pauli-$X$ operator $\sigma_i^x$, or vice versa. We make the first (second) choice for the red (green) sublattice in Fig.~\ref{cx}.

In the strong-$U$ limit, we can treat the Majorana hybridizations as a perturbation. The effective low-energy Hamiltonian emerges from local terms which leave the fermion parities of the Majorana  quartets unchanged, even though the latter are flipped by individual hybridization terms. The leading-order Hamiltonian involves 16-Majorana terms for the octahedra shown in Fig.~\ref{cx} (see Methods section below). Writing the Hamiltonian in the spin representation yields
\begin{align} 
& H=-\sum_{{\rm octahedra}}\left\{ \prod_{i \in \text{octa}^a} \sigma^x_i + \prod_{i \in \text{octa}^b} \sigma^z_i\right\}.
\label{pl}
\end{align}
Here $\text{octa}^a$ and $\text{octa}^b$ refer to the two types of octahedra in Fig.~\ref{cx} with red (green) sites at top and bottom and four green (red) sites in between. Thus, our construction exactly reproduces the planon-lineon model in Ref.~\cite{shirley2018fractional} whose elementary quasiparticles are lineons and planons with mobility restricted to the $z$-direction and the $xz$ ($yz$)- planes, respectively. 

{\em Experiment implementation.---}Our construction establishes a route towards building fracton codes from interacting Majoranas. To implement this construction, we must establish that one can realize the onsite interaction in Eq.~(\ref{int}).

To this end, we distribute the eight Majoranas of each site over two adjacent superconducting islands (SCI) as shown in Fig.~\ref{cx}. Each SCI could be made from two semiconductor quantum wires proximity coupled to the same superconductor. The proximity-coupled quantum wires effectively realize open Kitaev chains with two Majorana zero modes localized at their ends, so that there are a total of four Majoranas on each SCI. By virtue of their charging energy, each SCI can be tuned to have even fermion parity, effectively implementing the interaction terms $U(\gamma^1_i \gamma^3_i \gamma^8_i \gamma^5_i+\gamma^4_i \gamma^2_i \gamma^6_i \gamma^7_i)$ in Eq.~(\ref{int}) \cite{Plugge2017box,Karzig2017scale}. 

To generate the remaining four-Majorana interaction in Eq.~(\ref{int}), we turn on inter-island Majorana hybridization with amplitude $t$,
\begin{align} 
&H_{t}=it(\gamma^3 \gamma^4+\gamma^8 \gamma^7)
\end{align}
Note that at the same time, there is no hybridization between Majoranas $\gamma^1_i$ and $\gamma^2_i$ as well as $\gamma^5_i$ and $\gamma^6_i$. These inter-island hybridizations can in principle be implemented by direct tunnel coupling. Alternatively, and perhaps more flexibly, one can bridge between the two Majorana islands using a coherent link \cite{Karzig2017scale}. Such a link consists of an additional proximity-coupled quantum wire with its fermion parity fixed by charging energy. Its two Majorana end states would then be hybridized with the two Majorana end states of the Majorana islands which one wants to hybridize. Since the hybridization between the Majoranas on the coherent link and the islands can be realized through gate controlled tunnel junctions, the hybridization strength is tunable. We finally mention that the same hybridization via coherent link can be used to implement the Majorana hybridizations between different sites as given by Eq.~(\ref{H}). 

In the limit of a large charging energy, which fixes the fermion parities of the SCI, a single Majorana tunneling between the islands is suppressed and the lowest order processes involve pairs of Majorana tunneling terms. At leading order, the resulting Hamiltonian is,
\begin{align} 
&H_{\rm eff}=U(\gamma^1_i \gamma^3_i \gamma^8_i \gamma^5_i+\gamma^4_i \gamma^2_i \gamma^6_i \gamma^7_i)+
\frac{ct^2}{U}\gamma^3_i \gamma^4_i \gamma^7_i \gamma^8_i,
\end{align}
where $c$ is a number of order one. This produces an anisotropic version of the onsite interaction in Eq.~(\ref{int}). The anisotropy of the interaction coefficients does not modify the ground state manifold and thus the low-energy spin-$1/2$ degree of freedom. The additional weak hybridization $t'$ of Majoranas on nearest-neighbor sites as described by Eq.~(\ref{H}) generates the spin interactions on the octahedra. As a sufficient condition, the local spin-$1/2$ degrees of freedom remain intact in the limit $U>t>t'$ such that  ${t^2}/{U}>t'$, and the resulting effective Hamiltonian realizes the planon-lineon code at leading order. 

\begin{figure}[t]
  \centering
      \includegraphics[width=0.45\textwidth]{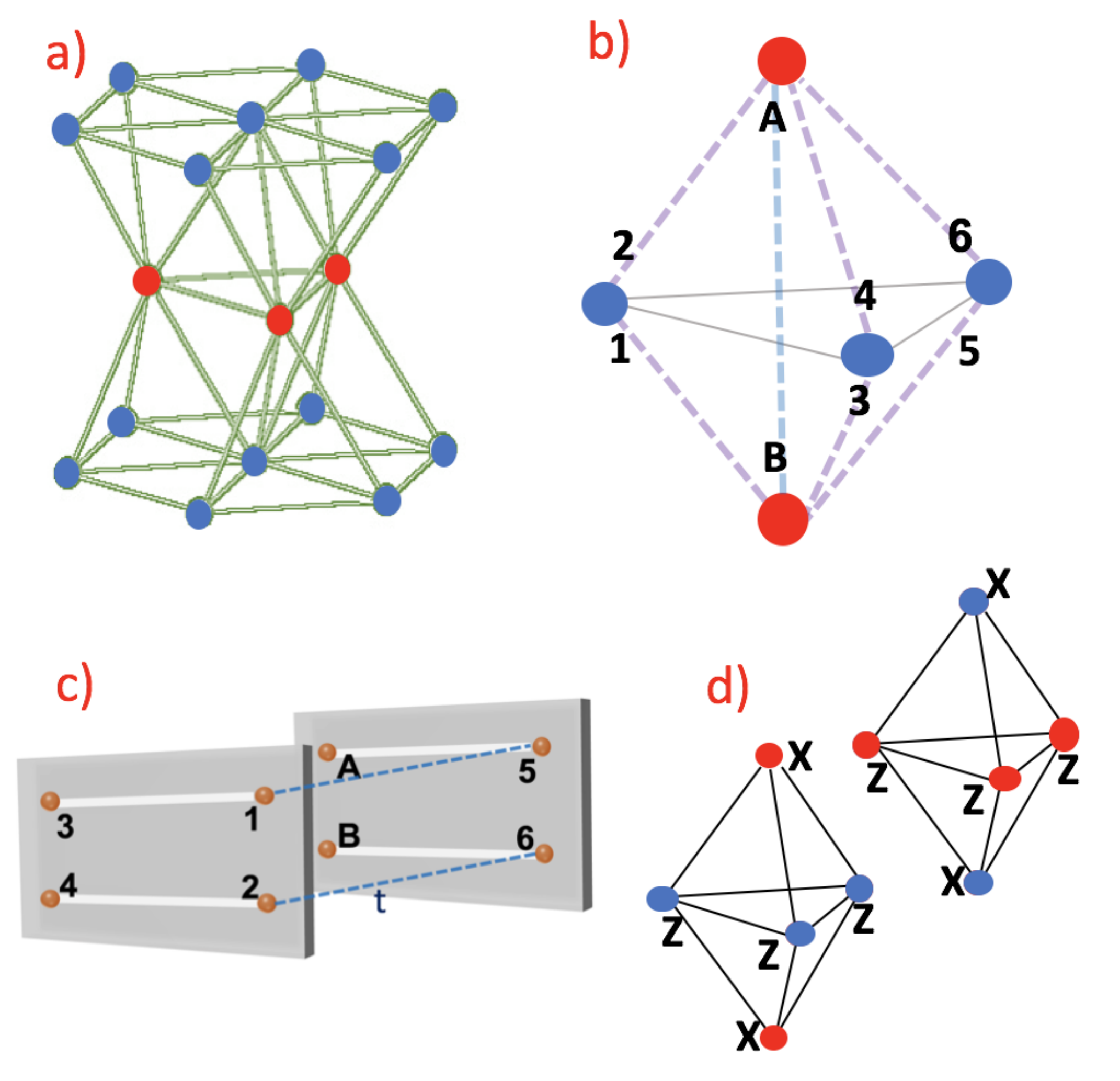}
  \caption{Majorana construction for the fractal Chamon code, a type-II fracton. a): Hexagonal close packed lattice with eight Majoranas per site with an elementary prism. b) The dashed purple lines illustrate the Majorana hybridizations for $\gamma^1,\ldots,\gamma^6$ and the dashed blue line the hybridizations for $\gamma^A,\gamma^B$. c) Onsite projection is implemented by placing Majorana wires on floating superconducting islands with inter-island tunneling and strong charging energy. d) The effective Hamiltonian after projection becomes a sum of five-spin stabilizers on the elementary prisms.} 
  \label{fractal}
\end{figure}

\begin{center}
\textbf{
The fractal Chamon code}
\end{center}

Following the spirit of this construction, one can realize many fracton stabilizer codes based on crossing Kitaev chains and strong onsite interactions. Specifically, we show now how a type-II fracton code -- the fractal Chamon code -- emerges in this manner. Unlike type-I fracton codes whose excitations are created by straight-line or planar membrane operators, the excitations of type-II fracton codes involve operators with fractal support \cite{Haah2011-ny,yoshida2013exotic}. In the case of the fractal Chamon code \cite{castelnovo2012topological}, excitations live at the corners of 2D Sierpinski triangles of side length $2^n$ \footnotemark[\value{footnote}]. These corner excitations are immobile in the $xy$ plane  unless one enlarges the Sierpinski triangle to side length $2^{n+1}$ which is inhibited by a large energy barrier.

The configuration of crossing Kitaev chains illustrated in Fig.~\ref{fractal} generates a hexagonal close-packed lattice, with each site containing eight Majoranas. Six of these labeled $\gamma^1, \ldots , \gamma^6$ are paired to Majoranas on nearest-neighbor sites in the same $xy$ plane, while the remaining two (labeled $\gamma^A$ and $\gamma^B$) are paired with Majoranas on neighboring sites along the $\pm z$ direction. 

Now consider the onsite interaction
\begin{align} 
&H=U (\gamma^1 \gamma^2 \gamma^4 \gamma^3+\gamma^1 \gamma^2 \gamma^6 \gamma^5+\gamma^5 \gamma^6 \gamma^B \gamma^A).
\label{fu}
\end{align}
In the strong-coupling limit, the eight Majoranas retain a spin-${1}/{2}$ degree of freedom. As for the planon-lineon model, we can define spin operators through $\sigma_x = \gamma_1 \gamma_3\gamma_5\gamma_B=\gamma_2\gamma_4 \gamma_6 \gamma_A$ and $\sigma_z=i\gamma_1\gamma_2 = i\gamma_3 \gamma_4 = i\gamma_5\gamma_6$. With these definitions, the low-energy Hamiltonian takes the form of the fractal Chamon code with five-qubit stabilizers defined on prisms as illustrated in Fig.~\ref{fractal} \cite{castelnovo2012topological}.

The elementary excitations include lineons with restricted mobility along the $z$-direction, generated by a line of $\sigma_z$ operators, in addition to the fractal excitations within the $xy$ planes which live at the corners of 2D Sierpinski triangles.  

The onsite interaction in Eq.~(\ref{fu}) can be implemented as for the planon-lineon code. The Majoranas involved in $\gamma^1 \gamma^2 \gamma^3 \gamma^4$ and $\gamma^A \gamma^B \gamma^5 \gamma^6$ are placed onto two floating SCI as shown in Fig.~\ref{fractal}. The charging energy of each SC island fixes their fermion parity. Hybridizations $it(\gamma^1_i \gamma^5_i+\gamma^2_i \gamma^6_i)$ effectively generate the remaining four-Majorana interaction $\gamma^1_i \gamma^5_i \gamma^2_i\gamma^6_i$, thereby exactly reproducing the interaction in Eq.~(\ref{fu}).

\begin{figure}[t]
  \centering
      \includegraphics[width=0.42\textwidth]{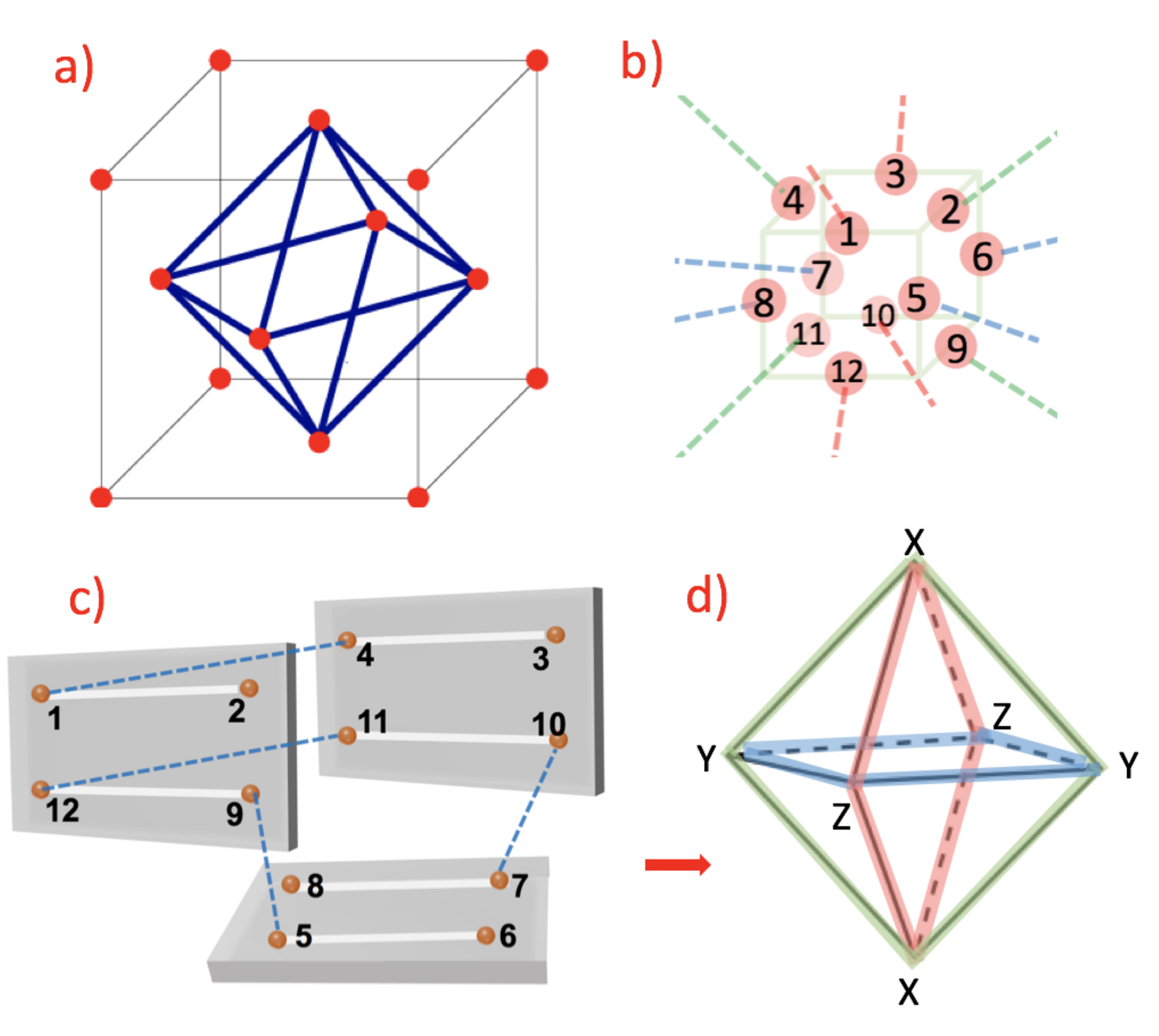}
  \caption{Majorana model for the octahedral Chamon code. 
 a) Face-centered cubic lattice with 12 Majoranas per site. b) Labeling and hybridization of Majoranas on one of the sites. c) Strong onsite interactions, realizable by this arrangement of floating superconducting islands, project each site into a spin-$1/2$ degree of freedom. d) Stabilizers of  the low-energy Hamiltonian live on the octahedra and the effective Hamiltonian becomes the octahedral Chamon code with fracton topological order.} 
  \label{chamon}
\end{figure}

\begin{center}
\textbf{
The octahedral Chamon code}
\end{center}

We now consider a system of hybridized Majoranas on a face-centered cubic lattice with 12 Majoranas on each site. Each of these Majoranas, placed on the edges of a cube, pairs with their partner on one of the nearest-neighbor sites as illustrated in Fig.~\ref{chamon}.

For each site, we place the 12 Majoranas onto three SCI as shown in Fig.~\ref{chamon}. The charging energy of the SCI fixes the fermion parities $\eta^1\eta^2 \eta^9 \eta^{12}=\eta^4 \eta^3 \eta^{10} \eta^{11}=\eta^7\eta^6 \eta^5 \eta^{8}=-1$. Weak tunneling between the different SCI islands,
 \begin{align} 
&H_{t'}=it(\eta^1 \eta^4+\eta^{12} \eta^{11}+\eta^9 \eta^5+\eta^{7} \eta^{10})
\label{typ}
\end{align}
generates the effective Majorana interactions $\eta^1 \eta^4\eta^{12} \eta^{11}$ and $i\eta^9 \eta^5 \eta^7 \eta^{10}  \eta^{12} \eta^{11}$. The full set of  interactions projects each site into a single spin-${1}/{2}$ degree of freedom and the corresponding Hamiltonian becomes the octahedral Chamon code \cite{Chamon2005-fc},
 \begin{align} 
&H= \sum_{\text{octahedra}}  \sigma^x_{r+e_x} \sigma^x_{r-e_x} \sigma^z_{r+e_z} \sigma^z_{r-e_z} \sigma^y_{r+e_y} \sigma^y_{r-e_y}.
\label{c2}
\end{align}
The six-spin terms associated with the octahedra originate from products of 12 Majorana pairs on the hinges (see SM for details). The Chamon code is a type-I fracton with quasiparticles of restricted mobility. The ground state degeneracy of this model depends on the greatest common divisor of the three spatial lengths.

\begin{figure}[t]
  \centering
      \includegraphics[width=0.4\textwidth]{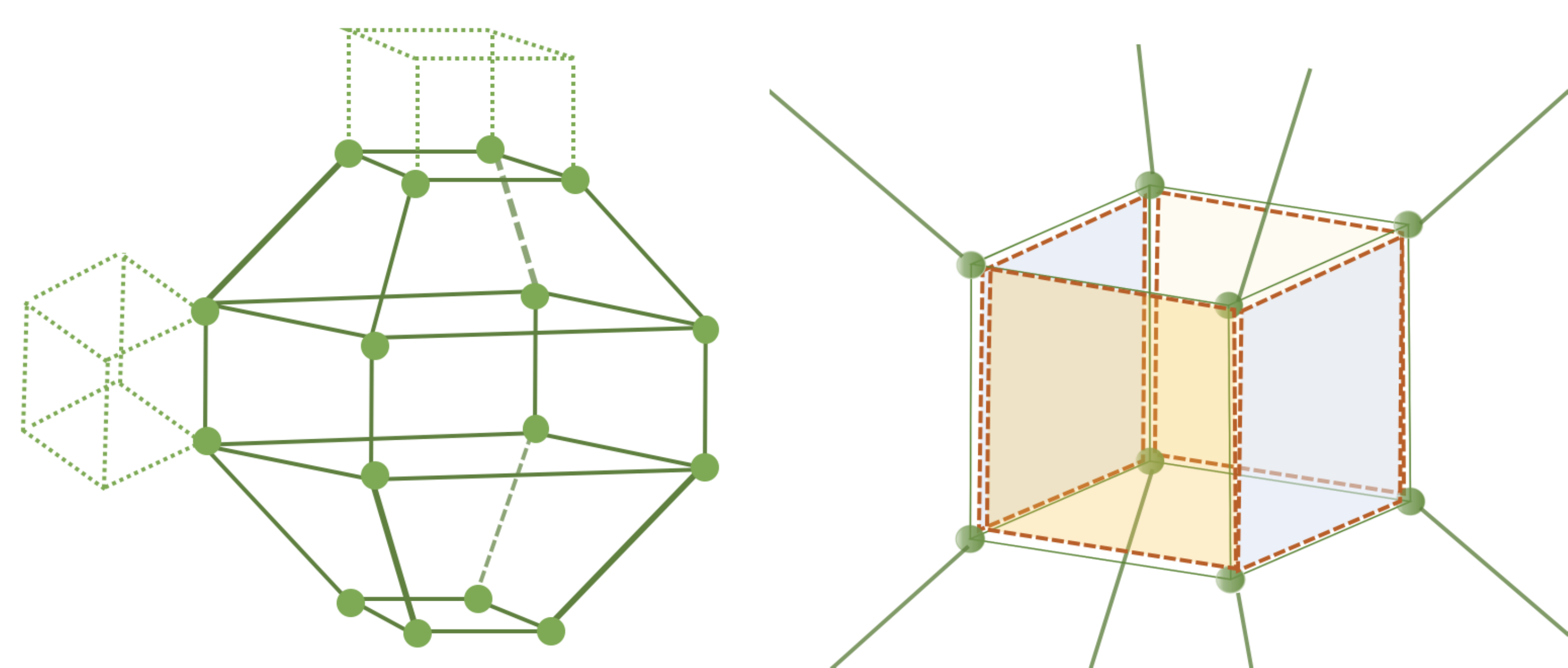}
  \caption{Left: The truncated octahedron shares a face with the top/bottom cubes and shares a hinge with the side cube. Right: The stabilizer is defined on the four side faces of the cube (yellow and blue), as well as the entire truncated-octahedron cell.} 
  \label{code}
\end{figure}

\begin{center}
\textbf{Fractons from
Majorana fermion codes}
\end{center}
Our fracton construction from interacting hybridized Majoranas suggests that many bosonic fracton models are equivalent to 3D Majorana fermion codes \cite{Bravyi2011-fl}. By enlarging each site of the bosonic model into a cell, the stabilizers of the Hamiltonian can be expressed as products over the Majoranas located on the corners of certain cells or plaquettes, provided that similar to color codes \cite{bombin2006topological}, the plaquettes and cells share an even number of Majoranas.

We make this argument explicit for the planon-lineon code in Fig.~\ref{cx}. We replace each site by a small cube, placing the eight Majoranas at the corners of the cube. The octahedra are then replaced by truncated octahedra as shown in Fig.~\ref{code} which share a face with cubes below or above and a hinge with a side cube. The spin Hamiltonian in Eq.~(\ref{int}) is replaced by the Majorana fermion code,
\begin{align} 
&H=-\sum \left\{\prod_{i \in \text{Octa}} \gamma_i+\prod_{i \in P^{xz}} \gamma_i+\prod_{i \in P^{yz}} \gamma_i
\right\}\label{surf}
\end{align}
The first term is a product over the 16 Majoranas at the corners of a truncated octahedron. The remaining terms involve products of four Majoranas on the side plaquettes ($xz$ or $yz$) of the cube. This Majorana fermion code defines a commuting projector Hamiltonian which exactly reproduces the ground-state manifold of the planon-lineon code. 

The same stratagem works for the other fracton codes which we discuss in this paper. While a wide variety of 3D Majorana fermion codes exhibiting $Z_2\times Z_2\times Z_2$ order have been proposed \cite{bravyi2010majorana}, our construction suggests that numerous fracton topological orders can also be represented by Majorana fermion codes akin to color codes. In particular, this reveals the equivalence between Majorana fracton codes and bosonic fracton codes.

{\em Fracton spin liquid.---}Now imagine we have four Majoranas per site, with each flavor forming a 
Majorana fermion code as given in Eq.~(\ref{surf}). By imposing a strong onsite inter-flavor interaction $U\gamma_1\gamma_2\gamma_3\gamma_{4}$, each site is constrained to even fermion parity and the corresponding low-energy Hilbert space is reduced to a spin-${1}/{2}$ degree of freedom. The resulting spin Hamiltonian is the SO(3) invariant fracton spin liquid
\begin{align} 
&H=-\sum_{a=x,y,z}\left\{\prod_{i \in \text{Octa}} \sigma^a_i+\prod_{i \in P^{xz}} \sigma^a_i+\prod_{i \in P^{yz}} \sigma^a_i\right\}.
\label{spinliquid}
\end{align}
One can also take $2N$ Majoranas per site and apply inter-flavor interactions to obtain an SO(2N-1) invariant Hamiltonian, yielding a hierarchy of spin liquids with fractonic behavior.

\begin{center}
\textbf{Discussion and experimental implementation}
\end{center}
We have demonstrated the emergence of
fracton phases of matter in systems of interacting Majorana band models. As discussed in Ref.~\cite{youcode}, the onsite interactions gauge the subsystem fermion parity symmetry and the resulting degrees of freedom obey a higher-rank $Z_2$ gauge theory describing a fracton phase of matter. On the theory side, our construction exposes the relation between fracton models and Majorana fermion codes. By extension, such Majorana fermion codes enable us to generate a hierarchy of fracton spin liquids with SO(N) symmetry. In addition to the examples discussed here explicitly, interacting Majorana wires can also realize additional models of fractons in particular and topological phases in general, including the checkerboard model proposed in Ref.~\cite{youcode}, the 3D toric code, the X-cube model, the cuboctahedron code, and the 2D fractal spin glass (see SM for details). 

On the experimental side, this provides a platform for exploring these novel strongly interacting phases. We conclude this paper by discussing opportunities for measuring and manipulating fracton phases of matter. 

\begin{figure}[t]
  \centering
      \includegraphics[width=0.47\textwidth]{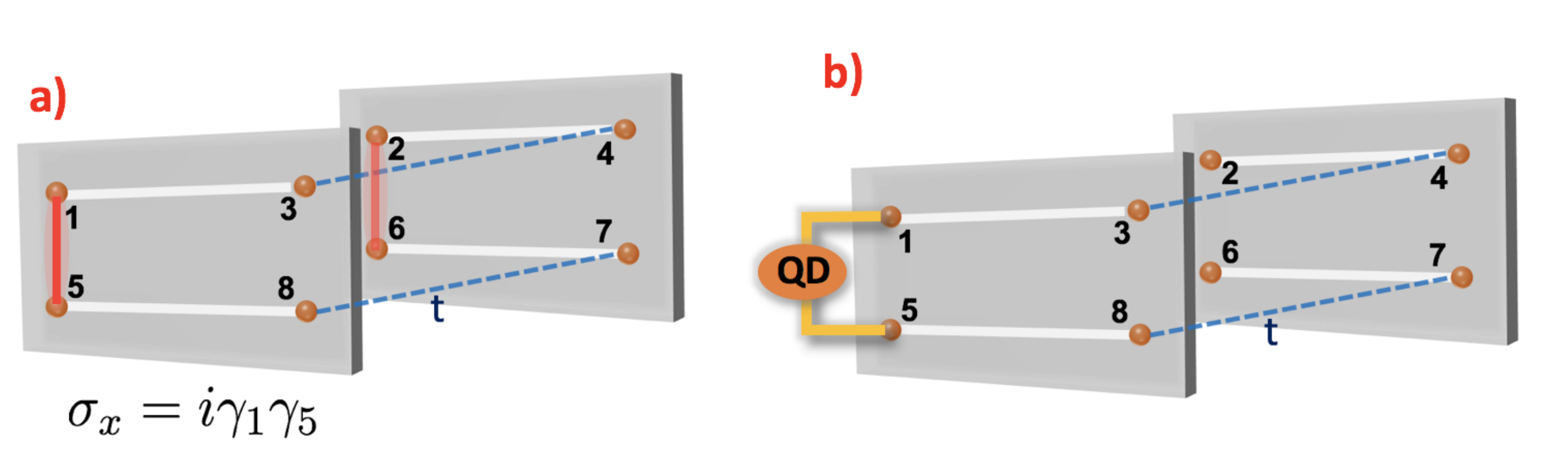}
  \caption{Measuring and manipulating fracton phases. a) Adding an onsite Majorana hybridization applies an effective Zeeman field to the low-energy spin degree of freedom associated with a site. The Zeeman field can be applied along the $x$, $y$, or $z$ directions, depending on which Majorana pair is hybridized. b) Measurement of the local spin for the case that the spin operator is represented by a Majorana bilinear. Coupling the Majoranas to a single level quantum dot shifts the quantum dot level energy by an amount which depends on the Majorana-parity associated with the bilinear. Thus, spectroscopy of the energy shift provides a measurement of the Majorana bilinear. Similar schemes can be implemented when the spin component is represented by a product of four Majoranas.} 
  \label{ex}
\end{figure}

So far, we have focused attention on realizing exactly-solvable stabilizer Hamiltonians of fracton codes. Sufficiently strong transverse fields applied to the local spin degrees of freedom drive the fracton phase into a confined phase \cite{Devakul2017-gg}. In our platform, such transverse fields are readily implemented by adding additional local Majorana hybridization terms as illustrated in Fig.~\ref{ex}. It would therefore be interesting to probe these phases and the intervening phase transition experimentally. While fracton phases lack a local order parameter, their phases and phase transitions are characterized by non-local string or membrane order parameters. For instance, the expectation values of certain planar Wilson loop operators may obey perimeter or area laws depending on the phase \cite{Devakul2017-gg}. Such expectation values can be read out by repeated preparation of a ground state and subsequent projective measurements of, say, the $\sigma_z$ components of all involved sites. Averaging the results for the string operators over the repeated measurements then provides access to the desired expectation value. Such projective measurements of local spins correspond to the measurement of local two- or four-Majorana parities for which several readout schemes have been proposed \cite{Plugge2017box,Karzig2017scale}. A promising scheme to measure, say, a two-Majorana parity couples a single-level quantum dot to the two Majoranas. The resulting energy shift of the quantum dot level will then depend on the Majorana parity. This is illustrated in Fig.~\ref{ex}. The same scheme can be extended to measure four-Majorana parities \cite{Karzig2017scale}. 

When applied only to subsets of sites, transverse fields can also be used to implement synthetic twist defects \cite{bombin2010topological}. For instance, adding intra-island Majorana hybridizations along a defect line, one can effectively create a twist defect of the fracton code which permutes different types of quasiparticles. This realizes quasiparticles of the 3D fracton phase which exhibit projective non-abelian statistics \cite{bombin2010topological}. The experimental tunability of the Majorana hybridization enables control over their location, an important prerequisite for probing their non-abelian braiding properties.

In addition to the static expectation values discussed above, one can also extract dynamic correlation functions of the fracton codes from linear-response measurements. Applying a time-dependent transverse field and reading out the spins at a later time provides access to spin-spin correlation functions of the fracton code. The dynamic properties of 3D fracton codes are particularly interesting due to their glassy dynamics which results from the restricted quasiparticle mobility. Moreover, disorder in the stabilizer flip energies or the transverse fields would would allow for studies of many body localization. Both kinds of disorder are easily tunable in this setting as inter-site hybridization and onsite stabilizer energies are controllable via gate-tuned coherent links.

\begin{center}
\textbf{Methods}
\end{center}
Traditionally, any spin model can be converted into a Majorana model by expressing the Pauli operators associated with each spin in terms of Majorana bilinears, made up from four Majorana operators $\gamma_1,\gamma_2,\gamma_3,\gamma_4$, and fixing the Majorana parity $\gamma_1 \gamma_2 \gamma_3 \gamma_4=-1$ on each site. 
In view of the Majorana anticommutation relations
\begin{equation}
  \{\gamma_i,\gamma_j\}=2\delta_{ij},
\end{equation}
the Majorana bilinears 
\begin{equation}
\sigma_x=i\gamma_2\gamma_3   \,\,\, , \,\,\, \sigma_y=i\gamma_1\gamma_3  \,\,\, , \,\,\,
\sigma_z=i\gamma_1\gamma_2
\end{equation}
satisfy the same algebra as the Pauli matrices. When applying this conventional fermionization to fracton codes, one again obtains complicated many-body interaction among fermion clusters. These are obviously difficult to implement, and the nontrivial challenge is thus to find a special Majorana representation which maps the spin cluster model to a Majorana model with local hybridization and merely onsite interaction implemented by charging energy. 
As we show above, this requires more involved Majorana representations of the spin degrees of freedom of the fracton codes, for instance involving also four-Majorana terms to represent Pauli matrices.

We now sketch the procedure how the spin cluster interactions corresponding to the stabilizers of the fracton models can be obtained from local interactions involving only small numbers of Majoranas. In the strong coupling limit, these interactions define a nearly degenerate low-energy subspace, with nontrivial Hamiltonian terms appearing only in high-order perturbation theory in the inter-site Majorana hybridizations. We sketch the underlying Brillouin-Wigner perturbation theory using the example of how to obtain the planon-lineon code from the strongly hybridized Majorana model in Eq.~(\ref{H}). Strong interactions in Eq.~(\ref{int}) enforce even fermion parity for each four-Majorana cluster. The effective Hamiltonian in this even-parity Hilbert space is obtained in seventh-order perturbation theory and takes the form
\begin{widetext}
\begin{align}
&H_{\rm eff}=\sum_j -O(\frac{t^8}{U^7})\gamma^3_{j+e_x+e_y} \gamma^8_{j+e_x+e_y} \gamma^2_{j-e_x-e_y} \gamma^6_{j-e_x-e_y} \gamma^1_{j-e_x+e_y} \gamma^5_{j-e_x+e_y} \gamma^4_{j+e_x-e_y} \gamma^7_{j+e_x-e_y} \nonumber\\
&\gamma^5_{j+e_z} \gamma^6_{j+e_z} \gamma^7_{j+e_z} \gamma^8_{j+e_z} 
\gamma^1_{j-e_z} \gamma^2_{j-e_z} \gamma^3_{j-e_z} \gamma^4_{j-e_z}
\end{align}
\end{widetext}
Defining spin operators through $\sigma_i^z=\gamma^1_i \gamma^2_i \gamma^4_i \gamma^3_i$ and the product of two Majoranas associated with any vertical edge as $\sigma_i^x$, the Hamiltonian becomes the planon-lineon code in Eq.~(\ref{pl}) with a six-spin cluster interaction on each octahedron. Any higher order perturbation terms are just products of the stabilizers in the Hamiltonian which do not change the ground state manifold.
Analogous arguments apply for all other fracton codes which we obtain from interacting Majorana models.

\begin{center}
\textbf{
Acknowledgments}
\end{center}

This work was supported in part by a PCTS fellowship (YY) and by CRC 183 of the Deutsche Forschungsgemeinschaft (FvO), and was initiated at the Aspen Center for Physics, which is supported by National Science Foundation grant PHY-1607611. FvO is grateful for sabbatical support from the IQIM, an NSF physics frontier center funded in part by the Moore Foundation. YY acknowledges the hospitality of IQIM-Caltech, where part of this work was completed. \textbf{Competing interests:} The authors declare that they have no competing interests. \textbf{Author contributions:} Both authors contributed to the theoretical results and experimental proposals presented here.
\textbf{Data availability:} All data needed to evaluate the conclusions in the paper are present in the paper and/or the Supplementary Materials. Additional data available from authors upon reasonable request.

\setcounter{figure}{0}
\setcounter{section}{0}
\setcounter{equation}{0}
\renewcommand{\theequation}{S\arabic{equation}}
\renewcommand{\thefigure}{S\arabic{figure}}

\onecolumngrid

\newcommand{\vsigma}{\mbox{\boldmath $\sigma$}}

\section*{\Large{Supplemental Material}}

\vspace{0.7cm}

\section{3D toric code and X-cube model}

Following the strategy in the main text, one can readily obtain the 3D toric code and the X-cube model from a similar Majorana construction. As pointed out in Ref.~\cite{Vijay2017-ey}, both models can be obtained via coupled layer constructions based on the 2D toric code. Moreover, the 2D toric code can be obtained via a Majorana network construction \cite{vijay2015majorana,youcode}. 

\begin{figure}[b]
  \centering
      \includegraphics[width=0.5\textwidth]{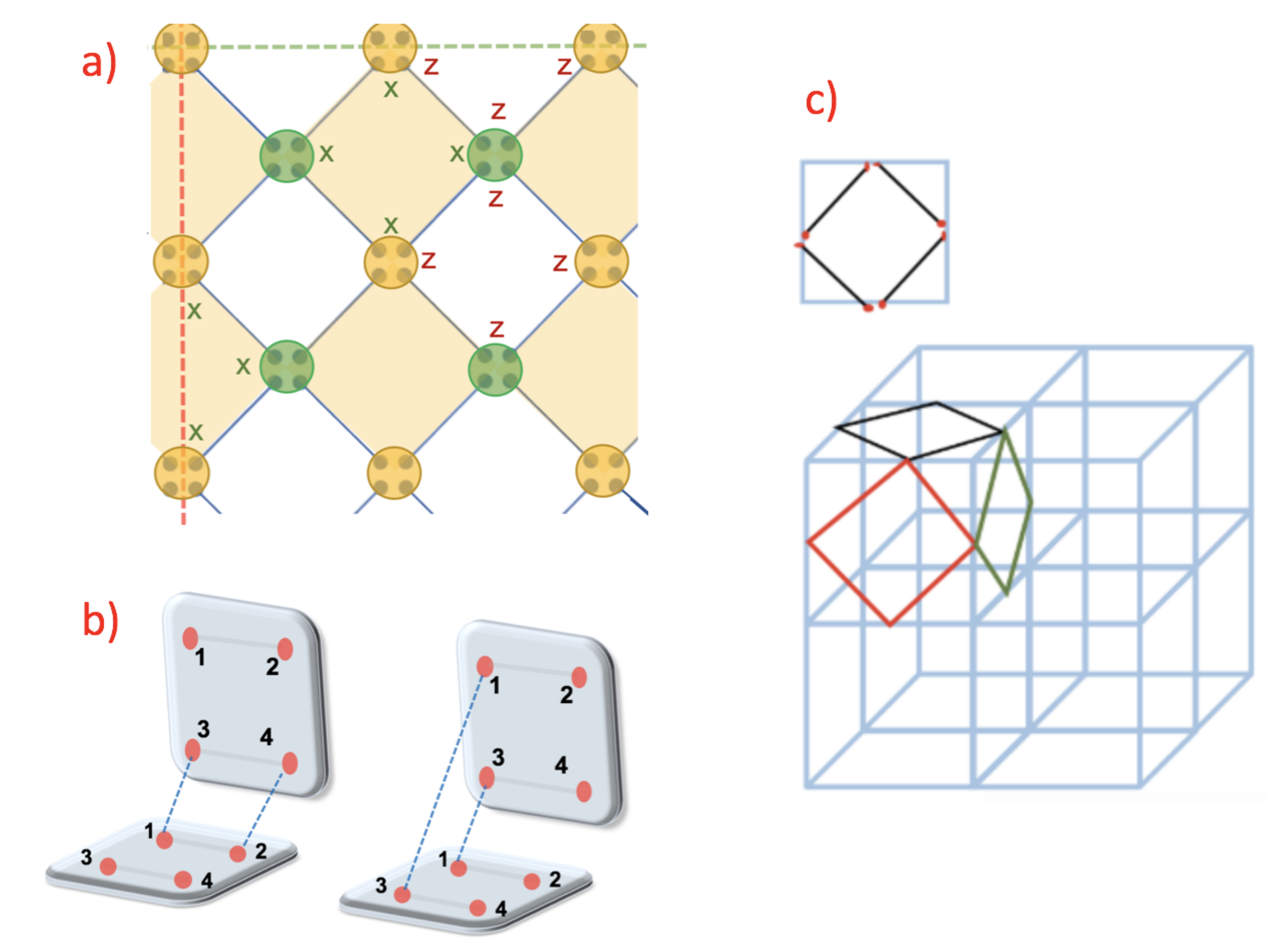}
  \caption{Majorana constructions for the X-cube model and 3D toric code. a) Toric code as a low-energy theory for the interacting Majorana band model. Each site supports a spin 1/2 and the plaquette operators involve product of $\sigma^z$ ($\sigma^x$) for white (yellow) plaquettes. b) Majorana interactions effectively implementing the coupled-layer construction. Each link of the cubic lattice has two SCI with four Majoranas per island which form 2D toric codes on the $ij$ layers. 
  Left: The interaction $H_A$ leads to the 3D X-cube model. Right: The interaction $H_B$ leads to the 3D toric code. c) Cubic lattice with two SCI on each link.}
  \label{sm1}
\end{figure}

We briefly review the Majorana construction for the 2D toric code. Consider a Majorana model on a square lattice with four Majoranas per site. Each Majorana hybridizes with its closest neighbor, as shown in Fig.~\ref{sm1} and described by the Hamiltonian 
\begin{align} 
H=-it \sum_j  (\gamma_j^1 \gamma_{j+e_r}^3+ \gamma_j^2 \gamma_{j+e_r'}^4)
\label{H1}
\end{align}
Here, the lattice sites are enumerated by $j$ and connected by lattice vectors $e_r$ and $e_r'$. Each site can be view as a SCI whose charging energy fixes the site's fermion parity,  $\gamma^1_j\gamma^2_j\gamma^3_j\gamma^4_j=-1$. The low-energy Hilbert space reduces to an effective spin-${1}/{2}$ degree of freedom on each site for which one can define Pauli operators (see, e.g., \cite{youcode}). Treating the Majorana hybridizations as perturbations, the leading-order Hamiltonian involves Majorana hopping terms around all elementary plaquettes. In terms of spin operators, the plaquette terms form a checkerboard pattern and become $\prod_{\square} \sigma^z_i \sigma^z_j \sigma^z_k \sigma^z_l$ and $\prod_{\square}\sigma^x_i \sigma^x_j \sigma^x_k \sigma^x_l$ for white and yellow plaquettes, respectively (see Fig.~\ref{sm1}). This is just the toric (or surface) code.

We can now implement the coupled layer construction by placing two SCI on each link of a cubic lattice. Each SCI hosts four Majoranas forming a 2D Toric code on each $ij$ layer. The model resembles decoupled 2D toric codes with each $i$-link holding qubits from the $ij$ and $ik$ layers as shown in Fig.~\ref{sm1}.

We can now add additional tunneling terms between Majoranas on same site but different SCI, see Fig.~\ref{sm1}. We consider two separate cases, described by the two Hamiltonians
\begin{align} 
&H_A=-it (\gamma_a^1 \gamma_b^3+ \gamma_a^2 \gamma_b^4)\nonumber\\
&H_B=-it (\gamma_a^1 \gamma_b^3+ \gamma_a^3 \gamma_b^1)
\end{align}
Here $a,b$ labels the two SCI on the same site. These tunneling terms, together with the fixed fermion parities of each SCI, generate effective onsite $XX$ or $ZZ$ interactions between spins from perpendicular layers, but on the same link. This couples the perpendicular 2d toric code layers. In the strong coupling limit, the interaction $H_B$ leads just to the 3D toric code model,
\begin{align} 
&H_{\rm eff}=\sum \left\{\prod_{i \in{\rm vertex}} \sigma^x_i+\prod_{i \in {\rm plaquette}}\sigma^z_i\right\}
\end{align}
The Hamiltonian involves six-spin vertex interacts involving $\sigma^z$ and four-spin plaquette interactions via $\sigma^x$. This model exhibits 3D $Z_2$ topological order with nontrivial particle loop braiding.

Likewise, the interaction $H_A$ leads to the 3D X-cube model
\begin{align} 
&H_{\rm eff}=\sum \left\{\prod_{ i\in v^{ij}} \sigma^x_i+\prod_{i \in {\rm cube}}\sigma^z_i\right\}.
\end{align}

\section{Triangle Ising model with fractal symmetry breaking}

Symmetries are indispensable for characterizing different phases of matter. Typically, one deals with global symmetries, whose operation acts extensively on an the volume of the system. Fractal subsystem symmetries, which act only on a subset of sites whose number scales with linear size L as $L^d$ with some fractal dimension $d$ have attracted much attention with the recent developments on fracton topological order. Systems with such symmetries appear most notably in the context of glassiness. Examples are the triangle or tetrahedral Ising models, whose Hamiltonians have fractal $Z_2$ symmetry when flipping spins on arbitrary Sierpinski triangles. We now show how to utilize a crossing Majorana network to generate the 2D triangle Ising model whose low energy ground states exhibit fractal symmetry breaking \cite{yoshida2013exotic}.

\begin{figure}[h]
  \centering
    \includegraphics[width=0.35\textwidth]{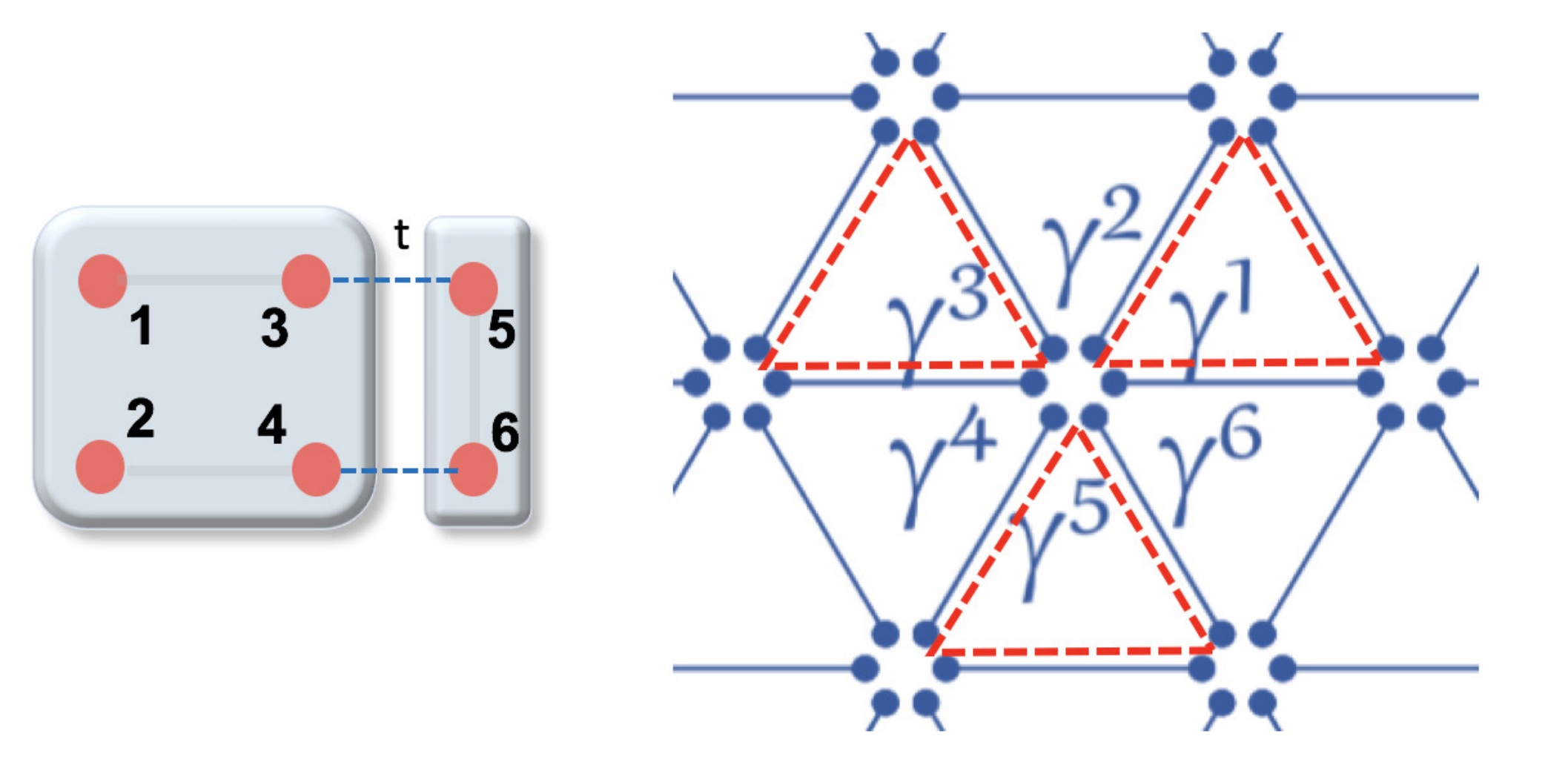} 
  \caption{Right: Crossing Kitaev wires yielding a triangle lattice with six Majoranas per site. Left: We place four Majoranas on one SCI and the other two on the other. The SCI holding four Majoranas is subject to a charging energy with fixed fermion parity. Majorana tunneling between the two SCI is indicated by dashed lines. This effectively produces the onsite interaction in Eq.~(\ref{oo}).} 
    \label{sm2}
\end{figure}

Consider a Majorana network on the honeycomb lattice with 
six Majoranas per site and each Majorana hybridized with a Majorana on a nearest-neighbor site as shown in Fig.~\ref{sm2}. This structure resembles three crossing Kitaev wires at angles $\theta={2N\pi}/{3}$. We impose the onsite interaction 
\begin{align} 
&H=-U(\eta^1_{i} \eta^2_{i} \eta^3_{i} \eta^4_{i}+ \eta^3_{i} \eta^4_{i} \eta^5_{i} \eta^6_{i}+\eta^1_{i} \eta^2_{i} \eta^5_{i} \eta^6_{i}).
\label{oo}
\end{align}
In the large-$U$ limit, the interaction enforces $\eta^1_{i} \eta^2_{i} \eta^3_{i} \eta^4_{i}= \eta^3_{i} \eta^4_{i} \eta^5_{i} \eta^6_{i}=\eta^1_{i} \eta^2_{i} \eta^5_{i} \eta^6_{i}=1$
As the third term is the product of the first two, these are two independent constraints and the six site Majoranas are projected into a spin-$1/2$ subspace. The resultant Hamiltonian involves the products of the three Majorana pairs in the upward triangle. Written in the spin basis, the Hamiltonian reduces to the triangle Ising model,
 \begin{align} 
&H=\sum_{\nabla} \prod_{i \in \nabla} \sigma^z_i
\end{align}
This Hamiltonian is invariant under fractal $Z_2$ transformations which flip the spins on arbitrary Sierpinski triangles. At zero temperature, the ground state breaks this fractal symmetry as characterized by the three-point correlator,
 \begin{align} 
&\langle \sigma^z_i(r_0) \sigma^z_i (r_0+a e_r) \sigma^z_i (r_0+a e_{r'})\rangle \neq 0, (a=2^{n} l_0).
\end{align}
This three-point correlation function does not vanish at long wavelengths. However, it fluctuates and becomes nonzero only when the three points hit the corners of a large Sierpinski triangle with distance $a=2^{n}l_0$.
This fractal structure breaks the discrete scale invariance. 

To implement the onsite fermion interaction in Eq.~(\ref{oo}), we place four and two Majoranas on one SC island each as shown in Fig.~\ref{sm2}. While the two-Majorana island is grounded, the left SC island with four Majoranas is floating so that its charging energy fixes the parity $\eta^1 \eta^2\eta^3 \eta^4=-1$. Weak tunneling $it(\eta^3\eta^5+\eta^4\eta^6)$ between the islands effectively creates the other four-Majorana interaction $\eta^3 \eta^4\eta^5 \eta^6$. 

\begin{figure}[t]
  \centering
      \includegraphics[width=0.45\textwidth]{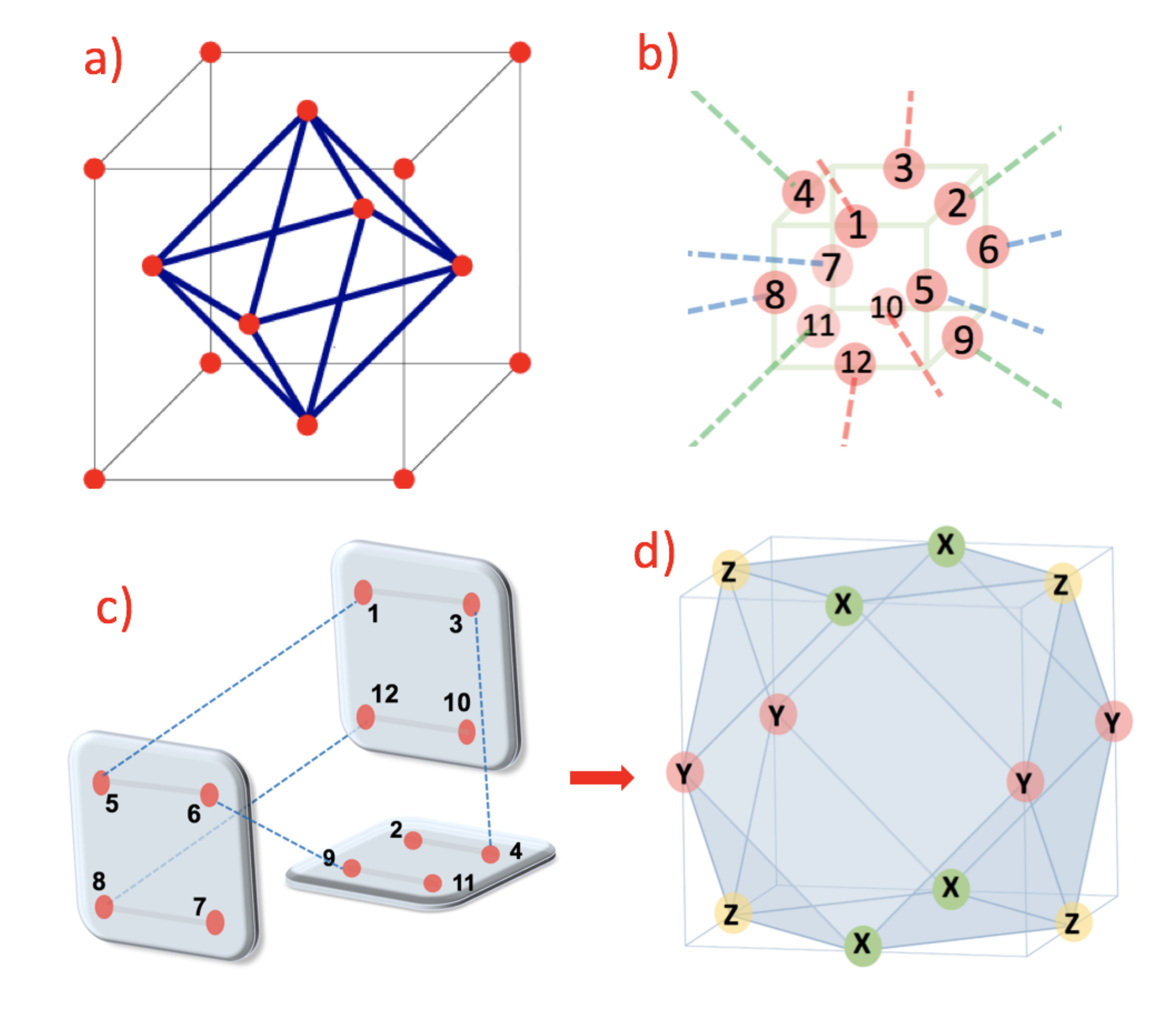}
  \caption{Majorana construction for the cuboctahedron code. a) Underlying face-centered cubic lattice. b) Interactions project each site into a spin-$1/2$ degree of freedom.  c) SCI design for implementing the onsite interactions. d) The effective spin Hamiltonian is a cuboctahedron fracton code. Each stabilizer is a 12-spin interaction on the cuboctahedron. The cuboctahedron is made of six corner-sharing plaquettes from the $xy$,$yz$,$xz$-planes.} 
  \label{sm3}
\end{figure}

\section{The cuboctahedron fracton code}

We now consider the same structure of crossing Kitaev chains as in Eq.~(\ref{typ}), yielding a face centered cubic lattice. At each site, there are twelve Majoranas, each of which pairs with a partner at a nearest neighbor site.


We place the 12 onsite Majorana onto three SCI as shown in Fig.~\ref{sm3}.  The charging energies of the three SC islands fixes the fermion parities $\eta^5\eta^6 \eta^7 \eta^{8}=\eta^1 \eta^3 \eta^{10} \eta^{12}=\eta^2\eta^4 \eta^{11} \eta^{9}=-1$. Turning on tunneling between the SC islands,
 \begin{align} 
&H_{t'}=it(\eta^5 \eta^1+\eta^{12} \eta^{8}+\eta^3 \eta^4+\eta^{9} \eta^{6})
\label{typ2}
\end{align}
generates effective Majorana interactions $\eta^1 \eta^5\eta^{8} \eta^{12}$ and $i\eta^3 \eta^4 \eta^8 \eta^{12}  \eta^{9} \eta^{6}$. When combined with the parity fixing for each SCI, these interactions project each site into a single spin-${1}/{2}$ degree of freedom and the Hamiltonian becomes the cuboctahedron code as shown in Fig.~\ref{sm3}. Each stabilizer is a 12-spin interaction on the cuboctahedron. The cuboctahedron is made of six corner-sharing plaquettes from the $xy$,$yz$,$xz$ planes. The 12-spin term on the cuboctahedron originates from the product of the 24 Majorana pairs on the hinges. This fracton code belongs to the family of type-I fracton codes whose quasiparticles have restricted mobility. Similar to the Chamon code, the number of ground-state degeneracy on a three-torus is dependent to the greatest common divisor of the system size.

The construction of the cuboctahedron code shares many similarities with the checkerboard code which we discussed in Eq.~\ref{c2}. The geometry of this Majorana network corresponds to four intersecting plaquettes sharing a corner at a site. The four-Majorana projections $\gamma^1 \gamma^3 \gamma^{10} \gamma^{12}, \gamma^5 \gamma^6 \gamma^{7} \gamma^{8}, \gamma^2 \gamma^9 \gamma^{11} \gamma^{4}$ produce three intersecting Wen-plaquette models on the $xy$, $yz$, and $xz$ planes labelled by spin qubits $(X^1,Z^1),(X^2,Z^2),(X^3,Z^3)$.
The remaining six Majoranas and four-Majorana interactions couple the three intersecting Wen-plaquette models via an anyon condensate. By imposing $Y_1 Y_2 Y_3=1$ and $Z_1 X_2=1$ on each site, the three spin qubits are reduced to one Pauli spin degree of freedom represented by $\sigma^x=X_1 Z_2,\sigma^y=X_2 Z_3,\sigma^z=X_3 Z_1$. The three corner-sharing intersecting Wen-plaquette models form an octahedron which exactly reproduces the Chamon code in Eq.~\ref{c2}. Likewise, if we impose $Y_1 Y_2 Y_3=1$ and $X_1 Z_2=1$ on each site, the three spin qubits are reduced to one Pauli spin degree of freedom represented by $\bar{\sigma}^x=Z_1 X_2,\bar{\sigma}^y=Z_2 X_3,\bar{\sigma}^z=Z_3 X_1$. This exactly reproduces the cuboctahedron structure which ties the six Wen-plaquette models on the six plaquette-faces of the cuboctahedron.
Our construction, en route, reveals the relation between 2D coupled toric code layer and 3D Chamon code from anyon condensate.

\section{SO(2N-1) Fracton spin liquid}
We place 2N Majoranas on each site, with each flavor forming a 
Majorana fermion code as given in Eq.~(\ref{surf}). By imposing a strong onsite inter-flavor interaction $U\gamma_1...\gamma_{2N}$, each site is constrained to even fermion parity and the the corresponding low-energy Hilbert space is reduced to spin-${(2N-1)}/{2}$ degree of freedom. One can use the Clifford algebra for Sp(2N-1) representation,
\begin{align} 
&\Gamma_a=i\gamma_1\gamma_{a+1} (a=1...2N-1)
\end{align}
The resulting spin Hamiltonian is the SO(2N-1) invariant fracton spin liquid
\begin{align} 
&H=-\sum_{a=1...2N-1}\left\{\prod_{i \in \text{Octa}} \Gamma^a_i+\prod_{i \in P^{xz}} \Gamma^a_i+\prod_{i \in P^{yz}} \Gamma^a_i\right\}
\end{align}

\end{document}